# Rechargeable redox flow batteries: Maximum current density with electrolyte flow reactant penetration in a serpentine flow structure

Xinyou Ke[1,2], Joseph M. Prahl[1], J. Iwan D. Alexander[1,3], Robert F. Savinell[2,4*]

[1]Department of Mechanical and Aerospace Engineering, Case Western Reserve University, Cleveland, Ohio 44106, United States

[2]Electrochemical Engineering and Energy Laboratory, Case Western Reserve University, Cleveland, Ohio 44106, United States

[3]School of Engineering, University of Alabama at Birmingham, Birmingham, Alabama 35294, United States

[4]Department of Chemical and Biomolecular Engineering, Case Western Reserve University, Cleveland, Ohio 44106, United States

*Corresponding author: rfs2@case.edu (Robert F. Savinell)

**Rechargeable redox flow batteries with serpentine flow field designs have been demonstrated to deliver higher current density and power density in medium and large-scale stationary energy storage applications. Nevertheless, the fundamental mechanisms involved with improved current density in flow batteries with flow field designs have not been understood. Here we report a maximum current density concept associated with stoichiometric availability of electrolyte reactant flow penetration through the porous electrode that can be achieved in a flow battery system with a "zero-gap" serpentine flow field architecture. This concept can explain a higher current density achieved within allowing reactions of all species soluble in the electrolyte. Further validations with experimental data are confirmed by an example of a vanadium flow battery with a serpentine flow structure over carbon paper electrode.**

Rechargeable redox flow batteries are considered as promising candidates for medium and large-scale stationary energy storage applications[1,2]. The electric energy stored by flow battery systems can be used to firm up intermittent renewable energy resources, and it can help to deliver consistent electricity to improve the stability of grids[3-6]. During recent decades, several types of flow batteries have emerged: all-vanadium[7,8], all-iron[9-11], zinc-polyiodide[12], semi-solid lithium ion[13,14], hydrogen-bromine[15,16], organic[17,18] and others. The merits of lower capital costs, eco-friendly, long-term life and higher electrochemical performance are desired[1-6]. The fundamental studies on electrodes[19], electrolytes[20], membrane[21,22] and cell design[4] are still on the way. Zawodzinski and Mench et al.[23,24] first reported a vanadium flow battery with a "zero-gap" serpentine flow field architecture. Higher current density and power density were observed in their cell with flow field design, which can drive down capital costs of the energy storage system. Since then, more work on flow batteries with flow field designs has been reported[25-30]. However, the detailed explanations involved with improved electrochemical performance in flow batteries with flow field designs are not yet given. This report presents an explanation of how the flow field designs improve current density in flow batteries and perhaps lead to a better understanding for further improvement. We have introduced the concept of maximum current density that can be achieved in rechargeable redox flow batteries with a single flow channel of a serpentine flow field design in our previous studies[27-30]. This maximum current density corresponds with 100% utilization of electrolyte flow reactants penetrating into the porous electrode as the electrolyte flow moves through the flow field.

In this report, we present several further important aspects beyond our previous work: (1) extend a two-dimensional model to a three-dimensional model for an actual serpentine flow field over porous electrode; (2) account for forced convective flow penetration into the porous electrode beneath the landings/ribs enhanced by pressure difference between the adjacent flow channels; (3) assess several models to correlate porosity and permeability of the porous electrode; (4) demonstrate reasonable agreement between our model with experimental data with a more realistic permeability of the porous electrode instead of an "adjusted" parameter and, (5) discuss the effects of electrode thickness, entrance volumetric flow rate and porosity/permeability of porous electrode on the maximum current density. The concept of maximum current density is validated by an example of experimental data from a vanadium flow battery with a "zero-gap" architecture serpentine flow field. This model explains observed limiting current densities and should provide some insights to understand higher electrochemical performance involved with forced convective mass transfer in redox flow batteries with flow field designs. This fundamental understanding may contribute to the electrochemical performance optimizations of rechargeable flow battery systems.

**Maximum Current Density Concept**

The classic understanding of a limiting current density is governed by a diffusion boundary layer near the electrode-electrolyte interface[31]. For electrolyte flowing through a porous electrode, the limiting current density can be written as:

$$i_{lim} = \frac{a t_p n F D c}{\delta_b} \quad (1)$$

Where, 'a' is the interfacial porous electrode surface area per unit porous electrode volume (~ 800 cm$^{-1}$, estimated by the Carla model[31] with a porosity of 0.8 and an average fiber diameter of 10 µm), '$t_p$' is the thickness of porous electrode (~ 0.04 cm), 'n' is the number of electrons transferred per mole of species reacted, 'F' is Faraday constant (96,485 C mol$^{-1}$), 'D' is the diffusivity (10$^{-6}$ cm$^2$ s$^{-1}$), 'c' is the bulk electrolyte concentration (0.001 mol cm$^{-3}$), and '$\delta_b$' is the thickness of diffusion boundary layer (~ 10 µm, estimated[4]). In the case where all electrolyte flows through a typical flow battery porous electrode, the liming current density is usually much larger than the stoichiometric limit, for example, ~ 3,200 mA cm$^{-2}$. This value is also much larger than the experimentally observed limiting current density (more than 400 mA cm$^{-2}$) in a vanadium flow battery with a serpentine flow field architecture design over multiple layers of carbon paper electrode[24]. Moreover, Newman et al.[32] also discussed a limiting current density achieved in forced convective electrolyte flow between two parallel plates where the one plate is the electrode with no flow penetration which is written as:

$$i_{lim} = 0.9783 \frac{nFDc}{L} \int_{o}^{L} \left(\frac{u_f}{hDX}\right)^{\frac{1}{3}} dX \tag{2}$$

In this equation, 'L' is the length of the electrode, '$u_f$' is the electrolyte flow velocity along the X-direction through the flow channel, 'h' is the distance between the electrode and backside of the flow channel. In the flow battery if the transport of reactant is controlled by the boundary layer between the electrolyte in the channel of the flow field and the outer electrode surface, then, Eq. (2) might be used to predict the limiting current density. Under an entrance volumetric flow rate of 0.333 cm$^3$ s$^{-1}$ (or 20 cm$^3$ min$^{-1}$), this limiting current density is estimated to be 50 – 100 mA cm$^{-2}$, which is significantly below the observed value in the experimental cells (more than 400 mA cm$^{-2}$) reported in a vanadium flow battery with a serpentine flow field structure[23,24]. The appearance of observed limiting current density may not be so much associated with interfacial diffusion, but more related to the availability reactant within the porous electrode. One can conclude that the convective electrolyte flow mass transport through the porous electrode from the flow field structure may play a significant role to support such a large current density. The mass balance and Faraday's law of electrolysis[27] yield (see details in supplementary information (SI), Section B: Methods)

$$i_{max} = \frac{nFc(Q_p)_{tot}}{A} \tag{3}$$

Where, '$(Q_p)_{tot}$' is the total volumetric flow of reactants penetration through the interface between the flow field and porous electrode layer and 'A' is the area of ion selective membrane. This expression assumes that all of the reactant penetrating into the porous electrode is consumed by Faradaic reaction. The flow penetration into the porous electrode underneath a serpentine flow channel is computed by a model using computational fluid dynamics simulation and then the maximum current density is estimated.

**Flow cell with flow structure**

The half-cell components of a vanadium flow battery with a serpentine flow channel over carbon paper electrode architecture are described in Fig. 1 (a) and (b). This half-cell structure was used as an example to model the observed limiting current density by estimating the amount of electrolyte flow reactant penetrating into the porous electrode from the serpentine flow channel. The half-cell structure consists of a current collector, a graphite plate engraved with a serpentine flow field, a gasket, a porous carbon electrode and an ion selective membrane. The electrolyte in the tank reservoir is circulated by a pump through the serpentine flow field and over the porous electrode. The cross-section views of the serpentine flow channel over electrode are shown in Fig. 1 (b): single flow passage and porous electrode (see view 1) and adjoin flow passages with landings/ribs over porous the electrode (see view 2). The three-dimensional geometry of the serpentine flow field with landings/ribs over the porous electrode used in this mathematical model is described in Fig. 1 (c) (three-dimensional XYZ view). A serpentine flow field over the porous electrode is simulated as shown in Fig. 1 (d) (two-dimensional XY view). The serpentine flow field consists of eleven flow passages (fp) and ten corner channels (cc). This geometry (and dimensions) represents a typical cell architecture reported in the literature[24]. The flow passages are designated as fp#1 through fp#11 and the corner channels are designated as

cc#1 through cc#10. The corresponding flow velocity component along the X, Y and Z directions is denoted as u, v and w. The details of the governing equations that describe flow motions within the serpentine flow field and porous electrode layer are provided in SI (Section B: Methods).

**Electrolyte flow streamline and interfacial flow velocity**

The streamline along the flow passages and corner channels over the porous electrode in the YZ plane is shown in Fig. 2 (a). It demonstrates that the electrolyte flow penetration is allowed from one flow passage through the porous electrode underneath a landing/rib to its adjoining flow passage. The streamlines under the landings/ribs are identified as "U" shapes with large curvatures. The interfacial flow velocity at the interface between the serpentine flow channel and porous electrode along the XY plane is shown in Fig. 2 (b). The interfacial flow velocity (w-component) reveals that electrolyte flow can be both into and out of the interface between the serpentine flow channel and porous electrode. More calculation results of flow velocity and pressure distributions at the middle thickness of the serpentine flow channel, interface between the serpentine flow channel and porous electrode and middle thickness of the porous electrode are presented in supplementary information.

**Volumetric flow penetration and maximum current density**

The electrolyte flow penetrating from the flow passages (through fp#1 to fp#11, see SI, Table S1) and corner channels (through cc#1 to cc#10, see SI: Table S1) into the porous electrode occurs through the interfaces from $\Omega_1$ to $\Omega_{21}$ (see SI: Table S1). The amount of electrolyte volumetric flow penetration into the porous electrode is estimated by an integration of w-component velocity along the –Z direction at interfaces from $\Omega_1$ to $\Omega_{21}$. The total volumetric flow penetration into the porous electrode is a sum of the volumetric flow penetrating into the interfaces from $\Omega_1$ to $\Omega_{21}$:

$$\left(Q_p\right)_{fp,total} = \sum_{\Omega_{j=1,odd}}^{\Omega_{j=21}} \iint |w_-| dXdY \tag{4}$$

$$\left(Q_p\right)_{cc,total} = \sum_{\Omega_{j=2,even}}^{\Omega_{j=20}} \iint |w_-| dXdY \tag{5}$$

$$\left(Q_p\right)_{tot} = \left(Q_p\right)_{fp,total} + \left(Q_p\right)_{cc,total} \tag{6}$$

Where, $|w_-|$ is an absolute value of w-component velocity along the –Z direction, j is the number of interface, $(Q_p)_{fp,total}$ is the total volumetric flow penetration into the interfaces between the flow passages and porous electrode, $(Q_p)_{cc,total}$ is the total volumetric flow penetration into the interfaces between the corner channels and porous electrode, and $(Q_p)_{total}$ is the total volumetric flow penetration into the porous electrode. Examples of the w-component velocity surface profiles along both the –Z and +Z directions at the interfaces from $\Omega_1$ to $\Omega_{21}$ are provided in SI. A positive (negative) value of w-component velocity means that flow reactant penetrates out of (into) the porous electrode. The volumetric flow penetration into the porous electrode at interfaces of flow passages/porous electrode and corner channels/porous electrode according to Eq. (4) to (6) are shown in Fig. 2 (c)-(f) for one layer and three layers of SGL 10AA carbon paper electrode. It can be seen that a sharp flow penetration into the porous electrode occurs at the interface between the #1 flow passage into the porous electrode. A possible explanation is that the electrolyte flow momentum at the entrance drives the electrolyte into the porous electrode. Although the flow penetrations at the interfaces between the corner channels and porous electrode (see Fig. 2 (d) and (f)) are much smaller compared with ones at the interfaces between the flow passages and porous electrode (see Fig. 2 (c) and (e)), the landings/ribs connected with the corner channels can enhance electrolyte fluid from one flow passage across the porous electrode underneath a landing/rib to its adjoining flow passage. Based on Eq. (3), the maximum current densities estimated from this model and experimental results for one layer and three layers of SGL 10 AA carbon paper electrode are compared in Fig. 3 (a) and (b). Under an entrance volumetric flow rate of 0.333 cm$^3$ s$^{-1}$ (20 cm$^3$ min$^{-1}$), the total volumetric flow rate through the porous electrode is computed to be 0.0238 cm$^3$ s$^{-1}$ (7.15% of entrance volumetric flow) and 0.0386 cm$^3$ s$^{-1}$ (11.6% of entrance volumetric flow) for a single layer and three layers of SGL 10 AA carbon paper electrode, respectively. The corresponding maximum current density predicted by Eq. (3) is 460 mA cm$^{-2}$ (vs. experimental result[24] of ~ 400 mA cm$^{-2}$) and 745 mA cm$^{-2}$ (vs. experimental result[24] of ~ 750 mA cm$^{-2}$ and experimental results[33] of 643 mA cm$^{-2}$ to 783 mA cm$^{-2}$). Further, the estimated maximum current density and amount of electrolyte flow penetration are sensitive to the permeability of the porous electrode. For example, a smaller permeability allows much smaller amount of electrolyte flow penetration and consequently permits a smaller maximum current density. In this report, a permeability of the porous electrode estimated as 10$^{-12}$ m$^2$ was used based on analysis of permeability correlations and experimental data as provided in supplementary information. In our earlier work[27-30], the permeability was only used as an adjustable parameter, and a high value of 10$^{-10}$ m$^2$ was employed in order to predict limiting current densities that agreed with reported experimental data in the literature. The corresponding effect of electrode thickness after compression and porosity after compression on maximum current density is shown in Fig. 3 (c) and (d), respectively. It can be seen that both electrode thickness and porosity have significant effects on maximum current density. Although thicker electrode allows a larger amount of electrolyte flow penetration, the increasing ability of electrolyte flow penetration decreases with the increased electrode thickness. It is worth mentioning that the ohmic loss cannot be ignored for thicker porous electrodes. The dashed line of ohmic limit in Fig.3 (c) is an estimate of the maximum current density as limited by ohmic loss if a 10% ohmic loss (20% round-trip) is acceptable for a typical vanadium electrolyte conductivity (100 mS cm$^{-1}$ to 200 mS cm$^{-1}$ for 1M V (II)/V(III), estimated[34,35]) and assuming equivalent porous electrode electronic conductivity. The volumetric mass transfer analysis may be a reasonable estimate if electrode performance in regions of Fig. 3 (c) left of the dashed line. Fig. 3 (d) shows the maximum current density estimated as a function of porosity of the porous electrode. It can be seen that larger porosity yields a larger maximum current density because of the greater electrolyte penetration into the electrode. On the other hand, larger porosity eventually will reduce the active surface area of the porous electrode, and thus kinetics and interfacial mass transfer will dominate the performance. Under this situation, the limiting current density estimated by Eq. (1) instead of maximum current density estimated by Eq. (3) will be a controlling mechanism limiting high current density electrochemical performance in rechargeable redox flow batteries with serpentine flow field designs.

**Conclusions**

In this report, we demonstrate a maximum current density concept associated with stoichiometric availability of electrolyte flow penetration through the porous electrode in rechargeable redox flow batteries with serpentine flow fields. The maximum current density controlled by the amount of reactant available is applicable to a thin porous electrode that have a small ohmic loss. Also, the interfacial mass transfer instead of reactant availability may limit the current density of the porous electrode having a larger porosity or smaller active surface area. The maximum current density approach presented here explains the experimentally observed high current density achieved in flow batteries with serpentine flow structures and thin carbon paper electrodes. This fundamental understanding should contribute to the optimization of electrochemical performance of rechargeable redox flow batteries with flow field designs.

**Methods**

This mathematical model is composed of a serpentine flow channel with eleven flow passages (fp) and ten corner channels (cc) over porous electrode in a vanadium flow battery with a "zero-gap" serpentine flow structure. The flow passages are designated as fp#1 through fp#11 and corner channels are designated as cc#1 through cc#10. The corresponding component of the flow velocity along the X, Y and Z directions is denoted by u, v and w. The major assumptions of this model are: (1) electrolyte flow is treated as steady, incompressible, Newtonian and laminar flow; (2) electrolyte flow through the porous electrode is considered as single phase and (3) no thermal effect is taken into account. The electrolyte flow dynamics in the flow field are governed by the Navier-Stokes equations. A macroscopic mathematical model is developed to capture flow physics in the porous electrode (see SI, Section B: Methods). The interface between the serpentine flow channel and porous electrode is divided into 11 sub-interfaces between the flow passages and porous electrode (denoted as $\Omega_j$, where j=1, 3, 5…21, see SI, Table S1) and 10 interfaces between the corner channels and porous electrode (denoted as $\Omega_j$, where j=2, 4, 6…20, see SI, Table S1). For the boundary conditions, the continuities of velocity and normal stress are applied at the interfaces from $\Omega_1$ to $\Omega_{21}$. Boundary conditions at the entrance are described by the following: (X=(wd$_p$-wd$_{fp}$)/2, 0.5l$_p$-5.5l$_{fp}$-5l$_{cc}$≤Y≤0.5l$_p$-4.5l$_{fp}$-5l$_{cc}$, t$_p$≤Z≤t$_p$+t$_f$), u$_{in}$=Q$_{in}$/(wd$_{fp}$t$_{fp}$) and at the outlet are described by the following: (X=(wd$_p$+wd$_{fp}$)/2, 0.5l$_p$+4.5l$_{fp}$+5l$_{cc}$≤Y≤0.5l$_p$+5.5l$_{fp}$+5l$_{cc}$, t$_p$≤Z≤t$_p$+t$_f$), p$_{out}$=0. All other boundary conditions are considered as "no slip" (u=0, v=0 and w=0). The serpentine flow channel and porous domains are refined with over three million unstructured triangular mesh. The type of advancing front triangular unstructured mesh is employed. This mathematical model is computed with a non-linear PARDISO algorithm solver embedded in COMSOL Multiphysics software together with self-written MATLAB programing codes and the relative error is $10^{-3}$. The details on the operation conditions, dimensions, properties of electrode and electrolytes are described in SI (see Table S4). The derivation of maximum current density model is given in SI (see Section B: Methods).

**Acknowledgments**

This work is supported by the all-iron flow battery project (DE-AR0000352) funded from ARPA-E program of Department of Energy (DOE) of the United States.

**Author contributions**

R.F.S. and X.K. proposed the idea. X.K. developed the model and used a computational approach to demonstrate the idea. X.K. wrote the manuscript. X.K., R.F.S., J.M.P. and J.I.D.A. revised the manuscript.

**Additional information**

Supplementary information is available for this manuscript. Correspondence and requests for materials should be addressed to R.F.S. (rfs2@case.edu).

**Competing financial interests**

The authors declare no competing financial interests.

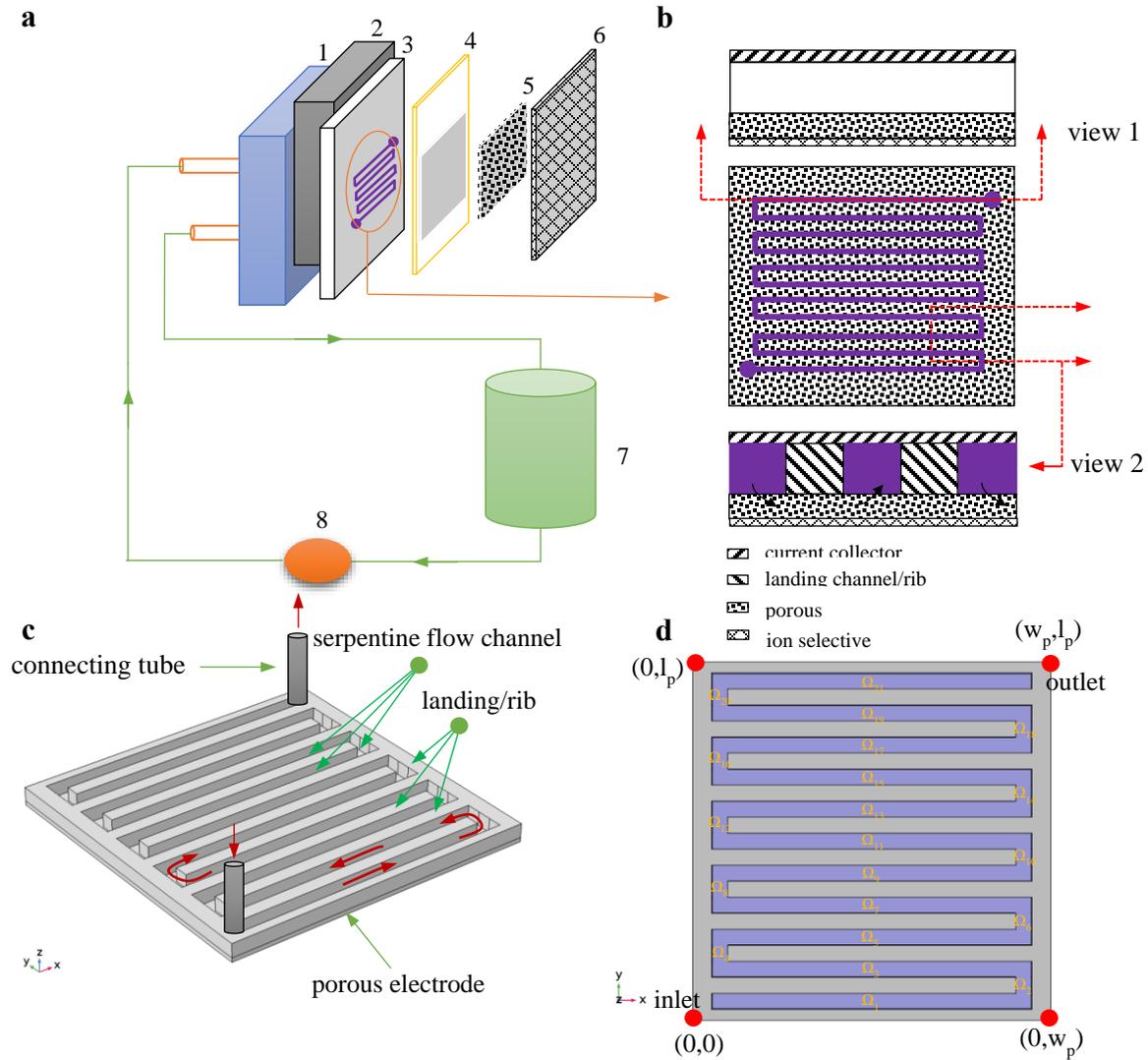

**Figure 1: Half-flow cell architecture. a,** 1-end plate; 2-current collector; 3-graphite plate engraved with a serpentine flow channel; 4-gasket; 5-porous electrode; 6-ion selective membrane; 7-eletrolyte tank and 8-pump. **b,** Cross-section views of a serpentine flow channel over porous electrode. **c,** Three-dimensional geometry of a serpentine flow field and landings/ribs over a porous electrode (XYZ view). **d,** Two-dimensional XY view of a serpentine flow channel over the porous electrode computed in the model, $\Omega_i$ (i=1, 3, 5, 7, 9, 11, 13, 15, 17, 19, 21) is defined as the i$^{th}$ interface between (i$^{th}$+1)/2 flow passage and porous electrode and $\Omega_j$ (j=2, 4, 6, 8, 10, 12, 14, 16, 18, 20) is defined as the j$^{th}$ interface between j$^{th}$/2 corner channel and porous electrode.

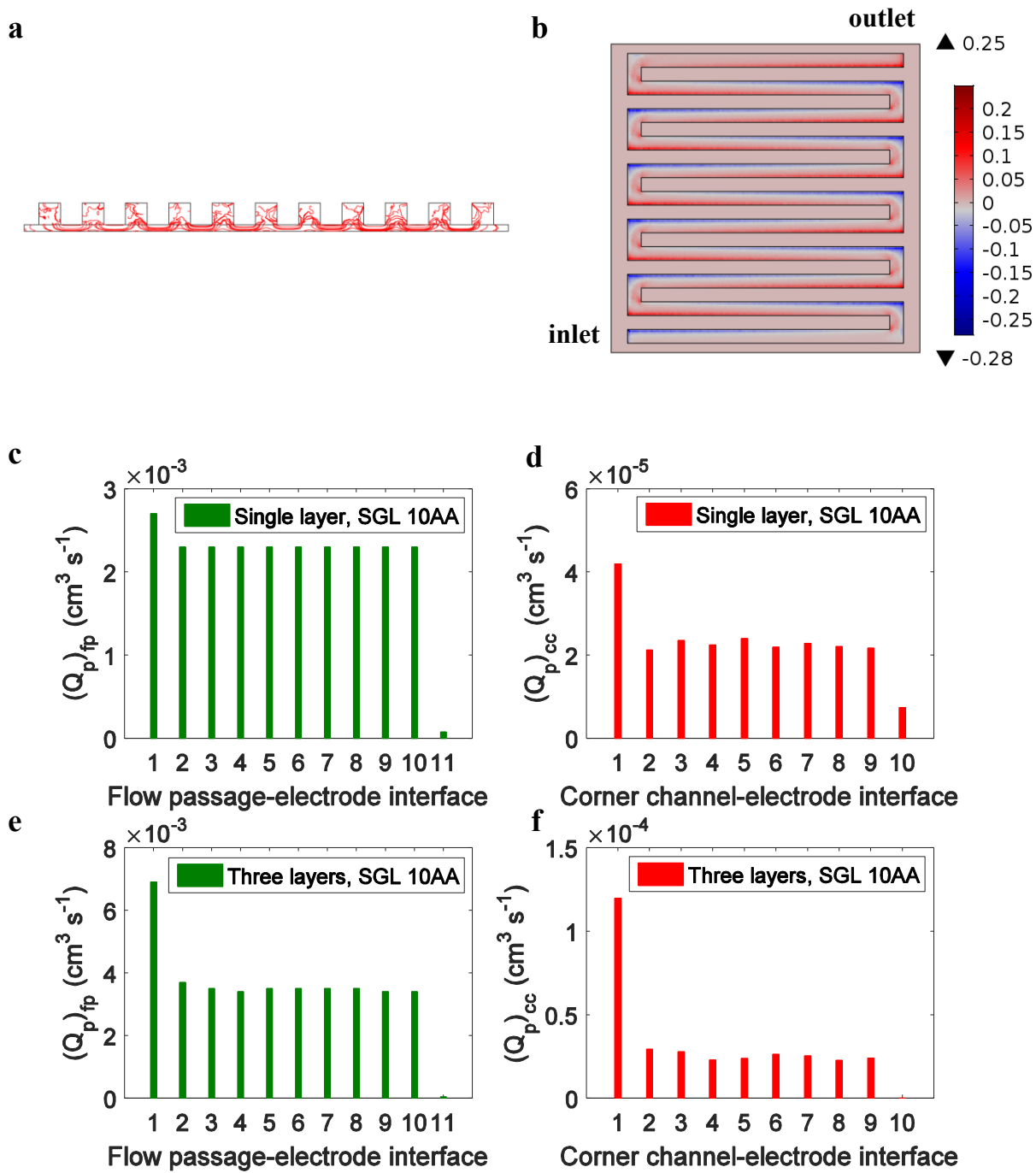

**Figure 2: Electrolyte flow streamlines, interfacial flow velocity and volumetric flow penetration. a,** Streamlines: (v, w) in the YZ plane, X=$w_p$/2 (middle width of the porous electrode). **b,** w-component velocity distribution at the interface between the serpentine flow channel and porous electrode, Z=$t_p$. **c,** Volumetric flow penetration through interfaces between the flow passages and porous electrode, one layer of SGL 10AA carbon paper electrode (308 µm compressed), unit: cm$^3$ s$^{-1}$. **d,** Volumetric flow penetration through interfaces between the corner channels and porous electrode, one layer of SGL 10AA carbon paper electrode, unit: cm$^3$ s$^{-1}$. **e,** Volumetric flow penetration through interfaces between the flow passages and porous electrode, three layers of SGL 10AA carbon paper electrode (908 µm compressed), unit: cm$^3$ s$^{-1}$. **f,** Volumetric flow penetration through interfaces between the corner channels and porous electrode, three layers of SGL 10AA carbon paper electrode, unit: cm$^3$ s$^{-1}$.

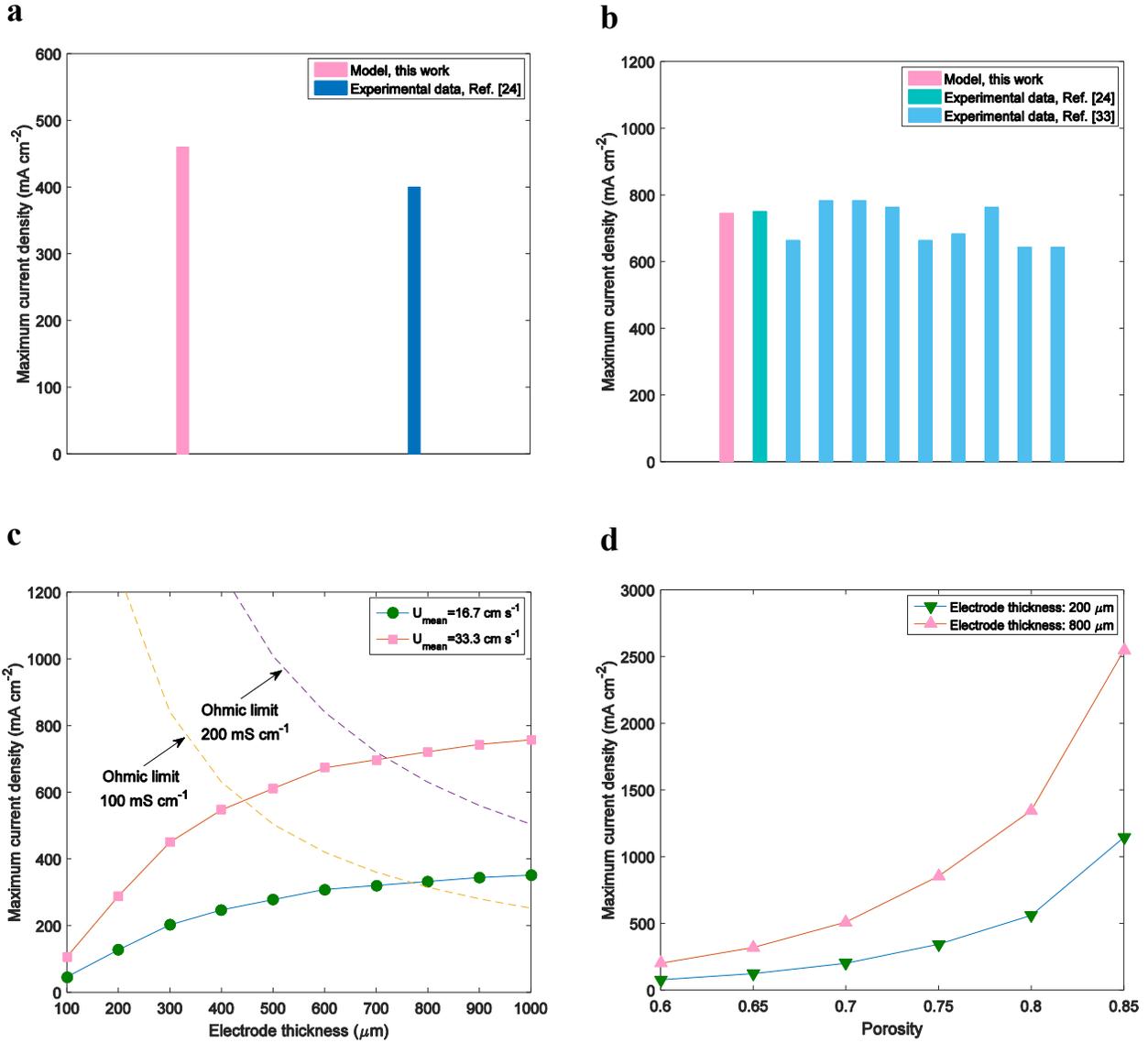

**Figure 3: Maximum current density associated with volumetric flow penetration. a,** Maximum current density (model vs. experimental data[24]), one layer of SGL 10AA carbon paper electrode, unit: mA cm$^{-2}$. **b,** Maximum current density (model vs. experimental data[24,33], three layers of SGL 10AA carbon paper electrode, unit: mA cm$^{-2}$. **c,** Effect of electrode thickness on maximum current density, mean entrance velocity of 16.7 cm s$^{-1}$ (entrance volumetric flow rate of 0.167 cm$^3$ s$^{-1}$) and 33.3 cm s$^{-1}$ (entrance volumetric flow rate of 0.333 cm$^3$ s$^{-1}$). The dash lines are the ohmic limits corresponding to an efficiency loss of 20% round-trip. **d,** Effect of porosity

on maximum current density, mean entrance velocity of 33.3 cm s$^{-1}$ (entrance volumetric flow rate of 0.333 cm$^3$ s$^{-1}$).

# Supplementary Information

# Rechargeable redox flow batteries: Maximum current density with electrolyte flow reactant penetration in a serpentine flow structure

Xinyou Ke[1,2], Joseph M. Prahl[1], J. Iwan D. Alexander[1,3], Robert F. Savinell[2,4*]

[1]Department of Mechanical and Aerospace Engineering, Case Western Reserve University, Cleveland, Ohio 44106, United States

[2]Electrochemical Engineering and Energy Laboratory, Case Western Reserve University, Cleveland, Ohio 44106, United States

[3]School of Engineering, University of Alabama at Birmingham, Birmingham, Alabama 35294, United States

[4]Department of Chemical and Biomolecular Engineering, Case Western Reserve University, Cleveland, Ohio 44106, United States

[*]Corresponding author: rfs2@case.edu (Robert F. Savinell)

## Section A: Nomenclature

A      area of ion selective membrane ($cm^2$)

BC      boundary condition

c      concentration (mol $cm^{-3}$)

cr      compression ratio

C      constant

F      Faraday constant (96,485 C $mol^{-1}$)

i      current density (A $cm^{-2}$)

k      permeability of the porous electrode ($m^2$)

L      length (cm)

n      number of electrons transferred in reactions

Q      volumetric flow rate ($cm^3$ $min^{-1}$ or $cm^3$ $s^{-1}$)

t      thickness (cm)

T      working temperature (K)

| | |
|---|---|
| u | X direction velocity (cm s$^{-1}$) |
| v | Y direction velocity (cm s$^{-1}$) |
| w | Z direction velocity (cm s$^{-1}$) |
| wd | width (cm) |
| X | X direction (cm) |
| Y | Y direction (cm) |
| Z | Z direction (cm) |

Greek symbols

| | |
|---|---|
| ε | porosity |
| μ | dynamic viscosity (Pa•s) |
| ρ | density of electrolyte flow (kg cm$^{-3}$) |
| Ω | interface |

Subscripts

| | |
|---|---|
| avg | average value |
| fp | flow passage |
| in | inlet |
| lc | landing channel |
| max | maximum |
| o | original |
| p | porous electrode |
| tot | total |

**Table S1:** Definitions of interfaces between the flow passages/corner channels and porous electrode.

| Symbols | Descriptions | Y coordinates |
|---|---|---|
| $\Omega_1$ | Interface between the fp#1 and porous electrode | $0.5l_p-5.5l_{fp}-5l_{cc} \leq Y \leq 0.5l_p-4.5l_{fp}-5l_{cc}$ |
| $\Omega_2$ | Interface between cc#1 and porous electrode | $0.5l_p-4.5l_{fp}-5l_{cc} < Y < 0.5l_p-4.5l_{fp}-4l_{cc}$ |
| $\Omega_3$ | Interface between fp#2 and porous electrode | $0.5l_p-4.5l_{fp}-4l_{cc} \leq Y \leq 0.5l_p-3.5l_{fp}-4l_{cc}$ |
| $\Omega_4$ | Interface between cc#2 and porous electrode | $0.5l_p-3.5l_{fp}-4l_{cc} < Y < 0.5l_p-3.5l_{fp}-3l_{cc}$ |
| $\Omega_5$ | Interface between fp#3 and porous electrode | $0.5l_p-3.5l_{fp}-3l_{cc} \leq Y \leq 0.5l_p-2.5l_{fp}-3l_{cc}$ |
| $\Omega_6$ | Interface between cc#3 and porous electrode | $0.5l_p-2.5l_{fp}-3l_{cc} < Y < 0.5l_p-2.5l_{fp}-2l_{cc}$ |
| $\Omega_7$ | Interface between fp#4 and porous electrode | $0.5l_p-2.5l_{fp}-2l_{cc} \leq Y \leq 0.5l_p-1.5l_{fp}-2l_{cc}$ |
| $\Omega_8$ | Interface between cc#4 and porous electrode | $0.5l_p-1.5l_{fp}-2l_{cc} < Y < 0.5l_p-1.5l_{fp}-l_{cc}$ |
| $\Omega_9$ | Interface between fp#5 and porous electrode | $0.5l_p-1.5l_{fp}-l_{cc} \leq Y \leq 0.5l_p-0.5l_{fp}-l_{cc}$ |
| $\Omega_{10}$ | Interface between cc#5 and porous electrode | $0.5l_p-0.5l_{fp}-l_{cc} < Y < 0.5l_p-0.5l_{fp}$ |
| $\Omega_{11}$ | Interface between fp#6 and porous electrode | $0.5l_p-0.5l_{fp} \leq Y \leq 0.5l_p+0.5l_{fp}$ |
| $\Omega_{12}$ | Interface between cc#6 and porous electrode | $0.5l_p+0.5l_{fp} < Y < 0.5l_p+0.5l_{fp}+l_{cc}$ |
| $\Omega_{13}$ | Interface between fp#7 and porous electrode | $0.5l_p+0.5l_{fp}+l_{cc} \leq Y \leq 0.5l_p+1.5l_{fp}+l_{cc}$ |
| $\Omega_{14}$ | Interface between cc#7 and porous electrode | $0.5l_p+1.5l_{fp}+l_{cc} < Y < 0.5l_p+1.5l_{fp}+2l_{cc}$ |
| $\Omega_{15}$ | Interface between fp#8 and porous electrode | $0.5l_p+1.5l_{fp}+2l_{cc} \leq Y \leq 0.5l_p+2.5l_{fp}+2l_{cc}$ |
| $\Omega_{16}$ | Interface between cc#8 and porous electrode | $0.5l_p+2.5l_{fp}+2l_{cc} < Y < 0.5l_p+2.5l_{fp}+3l_{cc}$ |
| $\Omega_{17}$ | Interface between fp#9 and porous electrode | $0.5l_p+2.5l_{fp}+3l_{cc} \leq Y \leq 0.5l_p+3.5l_{fp}+3l_{cc}$ |
| $\Omega_{18}$ | Interface between cc#9 and porous electrode | $0.5l_p+3.5l_{fp}+3l_{cc} < Y < 0.5l_p+3.5l_{fp}+4l_{cc}$ |
| $\Omega_{19}$ | Interface between fp#10 and porous electrode | $0.5l_p+3.5l_{fp}+4l_{cc} \leq Y \leq 0.5l_p+4.5l_{fp}+4l_{cc}$ |
| $\Omega_{20}$ | Interface between cc#10 and porous electrode | $0.5l_p+4.5l_{fp}+4l_{cc} < Y < 0.5l_p+4.5l_{fp}+5l_{cc}$ |
| $\Omega_{21}$ | Interface between fp#11 and porous electrode | $0.5l_p+4.5l_{fp}+5l_{cc} \leq Y \leq 0.5l_p+5.5l_{fp}+5l_{cc}$ |

**Section B: Methods**

*Discussions of porosity and permeability*

The literatures on porosity and permeability of carbon electrode materials, such as SGL 10 AA and Toray carbon paper electrodes are quite limited. One common experimental approach of measuring the permeability indirectly employs Darcy's law, and it correlates the pressure gradient with dynamic viscosity, bulk velocity, characteristic flow path length, and permeability for the single-phase flow

$$k = \frac{\mu \langle u_p \rangle L}{\Delta p} \tag{S1}$$

Where, k is the permeability, μ is the dynamic viscosity, $\langle u_p \rangle$ is the superficial velocity or bulk velocity, L is the characteristic length of flow path in the porous medium and Δp is the pressure difference. There are four models on estimating permeability summarized in Table S2.

**Table S2:** Summary of three models for predicting the permeability

| Permeability | Expressions | Descriptions | Sources |
|---|---|---|---|
| k | $\dfrac{d_f^2 \varepsilon^3}{C_{ck}(1-\varepsilon)^2}$ | Carman-Kozeny model, $C_{ck}$ is a fitting constant and varies with different porous materials | S1, S2 |
| | $\dfrac{d_f^2 \varepsilon (\varepsilon - \alpha_1)^{(\alpha_2+2)}}{2(\ln\varepsilon)^2 (1-\alpha_1)^{\alpha_2} (\alpha_2\varepsilon + \varepsilon - \alpha_1)^2}$ | Tomadakis-Sotirchos model for 3D random aligned fibers, $\alpha_1$=0.037, $\alpha_2$=0.661 | S3, S4 |
| | $0.012 d_f^2 \varepsilon \left( \dfrac{\pi^2}{16(1-\varepsilon)} - \dfrac{\pi}{2(1-\varepsilon)} + 1 \right) \left( 1 + 0.72 \dfrac{1-\varepsilon}{(\varepsilon - 0.11)^{0.54}} \right)$ | Tamayol-Bahrami model | S5 |
| | No explicit form | Derived from Lattice Boltzmann method, | S6 |

Gostick et al.[S5] pointed out that the typical permeability of GDLs (similar to the porous electrodes used in the flow battery technologies) used in the fuel cell designs ranges from $1\times10^{-11}$ m² to $5\times10^{-11}$ m². Weber et al.[S6] reported that the permeability is $2\times10^{-11}$ m² for the typical carbon felt or carbon fiber paper electrode. The permeability estimated by the Carman-Kozney model is $2.31\times10^{-10}$ m² with the following parameters: $\varepsilon=0.8$, $d_f=10$ μm and $C_{ck}=5.55$[S7-S9]. During the installation of flow cell components, the porous electrode is compressed and it is assumed that the shapes of porous fibers are not changed during the compression. The porosity of the porous electrode can be estimated after compression

$$\varepsilon_{p,c} = 1 - \frac{(1-\varepsilon_{p,o})t_{p,o}}{t_{p,c}} \tag{S2}$$

$$cr = 1 - \frac{t_{p,c}}{t_{p,o}} \tag{S3}$$

Where, $t_{p,o}$ is the original thickness of the porous electrode, $t_{p,c}$ is the thickness of the porous electrode after compression, $\varepsilon_{p,o}$ is the original porosity, $\varepsilon_{p,c}$ is the porosity after compression and cr is compression ratio. To be a physically meaningful quantity, $\varepsilon_{p,c}$ must satisfy the following inequality

$$\frac{t_{p,o}}{t_{p,c}} > 1 - \varepsilon_{p,o} \tag{S4}$$

The relationship between porosity after compression and ratio of the thickness after compression over the original thickness under several different original porosities is shown in Figure S1 (a).

The permeability is reduced when the porous electrode is compressed during the assembly of the flow cell. A graph of predicted permeability as a function porosity from several reported models is shown in Figure S1 (b). It can be seen that the Carman-Kozeny model[S1,S2] and Tomadakis model[S3,S4] overestimate the experimental value[S8] of permeability for the SGL 10 AA carbon paper electrode. It seems that Tamayol-Bahrami model[S5] and Doormaal-Pharoah model[S6] match better with experimental data for SGL 10AA carbon paper electrode[S10]. The average permeability of porous electrode with different porosity is calculated and summarized in Table S3. The corresponding estimated porosity and compressed thickness of SGL 10AA carbon paper electrode is estimated to be 0.734 by Eq. (S2) with an original porosity of 0.8 and 308 μm with a compression ratio of ~25%. The predicted average permeability is $5.6 \times 10^{-12}$ m$^2$ for the SGL 10AA carbon paper electrode.

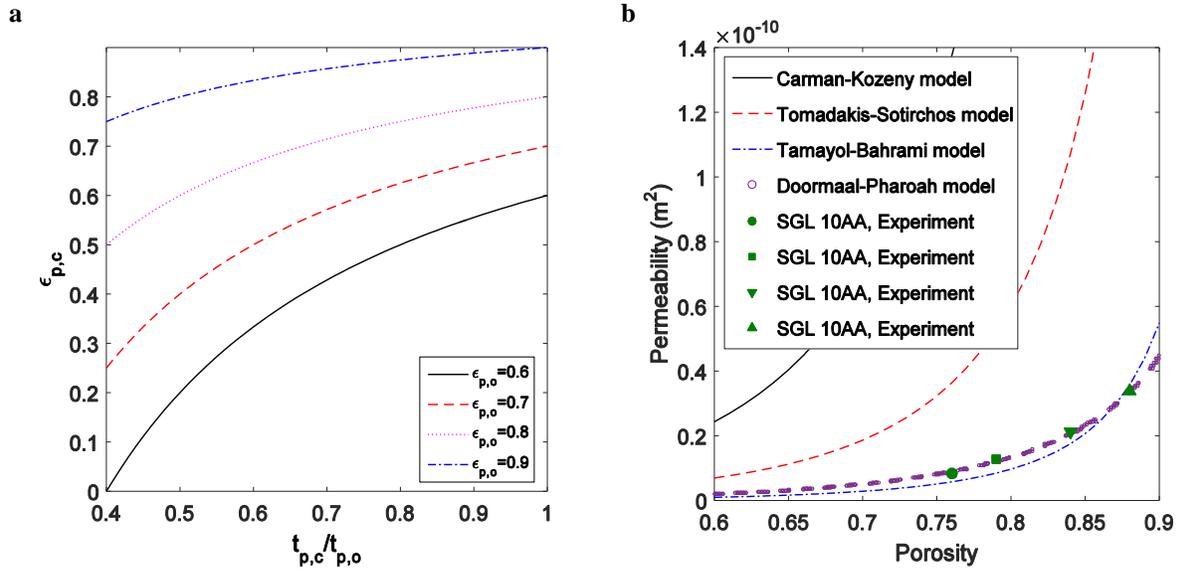

**Figure S1: a,** The relationship between $\varepsilon_{p,c}$ and $t_{p,c}/t_{p,o}$ under four different $\varepsilon_{p,o}$: 0.6, 0.7, 0.8 and 0.9. **b,** Estimations of permeability from several models. Carman-Kozeny model[S1,S2], Tomadakis model[S4], Tamayol-Bahrami model[S5] and Doormaal-Pharoah model[S6] vs. experimental data for SGL 10AA carbon paper electrode[S10]

**Table S3:** Estimations of porosity and permeability based on Tamayol-Bahrami model[S5] and Doormaal-Pharoah model[S6]

| $\varepsilon$ | k (m$^2$) Tamayol-Bahrami model | k (m$^2$) Doormaal-Pharoah model | k$_{avg}$ (m$^2$) |
|---|---|---|---|
| 0.6  | 2.0×10$^{-12}$ | 9.5×10$^{-13}$ | 1.5×10$^{-12}$ |
| 0.65 | 3.1×10$^{-12}$ | 1.6×10$^{-12}$ | 2.4×10$^{-12}$ |
| 0.7  | 5.0×10$^{-12}$ | 2.8×10$^{-12}$ | 3.9×10$^{-12}$ |
| 0.75 | 8.2×10$^{-12}$ | 5.1×10$^{-12}$ | 6.7×10$^{-12}$ |
| 0.8  | 1.3×10$^{-11}$ | 9.7×10$^{-12}$ | 1.1×10$^{-11}$ |
| 0.85 | 2.4×10$^{-11}$ | 2.1×10$^{-11}$ | 2.3×10$^{-11}$ |

*Parameters*

The dimensions, operation conditions, properties of electrolyte and electrode are described in Table S4. The length, width and thickness of the flow passage are typically corresponding to be ~2 cm, 0.1 cm and 0.1 cm in a typical 5 cm$^2$ laboratory flow cell. Here, the ratio of the flow passage width to the landing channel width is 1. The thickness of the uncompressed single layer of the 10 AA SGL and Toray carbon paper electrode is ~ 410 μm (308 μm compressed) and ~ 200 μm, respectively. The typical compress ratio is ~ 25% during flow cell assembly. The compress process will change the dimensions and properties of carbon paper electrodes.

**Table S4:** Dimensions, operation conditions, properties of electrolyte and porous electrode

| Symbols | Descriptions | Value | Units | Sources |
|---|---|---|---|---|
| $l_{fp}$ | Length of the flow passages | 2 | cm | Measured |
| $wd_{fp}$ | Width of the flow passages | 0.1 | cm | Measured |
| $t_{fp}$ | Thickness of the flow passages | 0.1 | cm | Measured |
| $l_{lc}$ | Length of the landing channels | 0.1 | cm | Measured |
| $wd_{lc}$ | Width of the landing channels | 0.1 | cm | Measured |
| $t_{lc}$ | Thickness of the landing channels | 0.1 | cm | Measured |
| $l_p$ | Length of the porous electrode | 2.24 | cm | Measured |
| $wd_p$ | Width of the porous electrode | 2.24 | cm | Measured |
| rc | Compression ratio | 25% | - | S11 |
| $t_{p,0}$ | Thickness of the porous electrode, original | 410 | μm | S11 |
| $t_{p,c}$ | Thickness of the porous electrode, compression | 308 | μm | Estimated by Eq. (S2) |
| $\varepsilon_0$ | Porosity, original | 0.8 | - | S11 |
| $\varepsilon_c$ | Porosity, compressed | 0.734 | - | Estimated by Eq. (S2) |
| k | Permeability | $5.6 \times 10^{-12}$ | $m^2$ | Estimated, Table S3 |
| ρ | Density of electrolyte flow | 1.35 | g cm$^{-3}$ | |
| μ | Dynamic viscosity of electrolyte flow | $4.93 \times 10^{-3}$ | Pa·s | |
| T | Working temperature | 298 | K | S7 |
| c | Ion concentration | 0.001 | mol cm$^{-3}$ | |
| $Q_{in}$ | Entrance volumetric flow rate | 20 | cm$^3$ min$^{-1}$ | |
| n | Number of transferring electrons in reactions | 1 | - | |
| F | Faraday constant | 96,485 | C mol$^{-1}$ | - |

*Macroscopic Mathematical Model*

The flow dynamics in the flow channel are governed by the mass conservation or continuity (Eq. (S5)) and Naveir-Stokes motion (Eqs. (S6)-(S8)) as the electrolyte flow is steady, incompressible, laminar and Newtonian flow regime along the X, Y and Z directions

$$\frac{\partial u_f}{\partial X} + \frac{\partial v_f}{\partial Y} + \frac{\partial \omega_f}{\partial Z} = 0 \tag{S5}$$

$$\rho u_f \cdot \frac{\partial u_f}{\partial X} + \rho v_f \cdot \frac{\partial u_f}{\partial Y} + \rho w_f \cdot \frac{\partial u_f}{\partial Z} = -\frac{\partial p_f}{\partial X} + \mu \left( \frac{\partial^2 u_f}{\partial X^2} + \frac{\partial^2 u_f}{\partial Y^2} + \frac{\partial^2 u_f}{\partial Z^2} \right) \tag{S6}$$

$$\rho u_f \cdot \frac{\partial v_f}{\partial X} + \rho v_f \cdot \frac{\partial v_f}{\partial Y} + \rho w_f \cdot \frac{\partial v_f}{\partial Z} = -\frac{\partial p_f}{\partial Y} + \mu \left( \frac{\partial^2 v_f}{\partial X^2} + \frac{\partial^2 v_f}{\partial Y^2} + \frac{\partial^2 v_f}{\partial Z^2} \right) \tag{S7}$$

$$\rho u_f \cdot \frac{\partial w_f}{\partial X} + \rho v_f \cdot \frac{\partial w_f}{\partial Y} + \rho w_f \cdot \frac{\partial w_f}{\partial Z} = -\frac{\partial p_f}{\partial Z} + \mu \left( \frac{\partial^2 w_f}{\partial X^2} + \frac{\partial^2 w_f}{\partial Y^2} + \frac{\partial^2 w_f}{\partial Z^2} \right) - \rho g \tag{S8}$$

The macroscopic mathematical model of describing flow dynamics in the porous electrode are derived by the transport theorem and averaging volume method and is largely based on previous studies of Whitaker[S8,S12,S13], Gary et al.[S14], Howes et al.[S15], Ochoa-Tapia et al.[S16,S17], Goyeau et al.[S18] and Bars et al.[S19]. The γ-σ phase system of porous electrode is shown in Figure S3. γ and σ phases are defined as electrolyte fluid and porous solid, respectively.

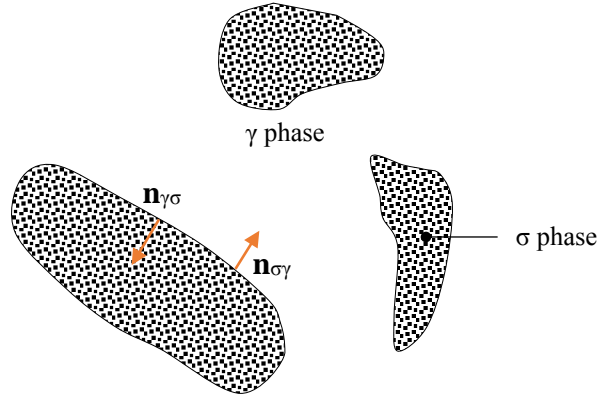

**Figure S2:** Normal unit vector $n_{\gamma\sigma}$ and $n_{\sigma\gamma}$ in the γ-σ phase system of the porous domain

$\mathbf{u}_{\gamma\sigma}$ is the velocity vector at the interface between γ phase and σ phase. The quantity dS denotes the interfacial boundary area between γ phase and σ phase, $\mathbf{n}_{\gamma\sigma}$ denotes the unit normal vector pointing from γ phase to σ phase. Similarly, $\mathbf{n}_{\sigma\gamma}$ points from the σ phase to the γ phase and $\mathbf{n}_{\sigma\gamma}$ is identical to -$\mathbf{n}_{\gamma\sigma}$. The transport theorem and averaging volume method yield

$$V_{\gamma\sigma} = V_{\gamma} + V_{\sigma} \tag{S9}$$

$$\langle \psi \rangle = \frac{1}{V_{\gamma\sigma}} \int \psi dV \tag{S10}$$

$$\langle \psi_{\gamma} \rangle = \frac{1}{V_{\gamma\sigma}} \int \psi_{\gamma} dV \tag{S11}$$

$$\langle \psi_{\gamma} \rangle^{\gamma} = \frac{1}{V_{\gamma}} \int \psi_{\gamma} dV \tag{S12}$$

$$\langle \psi_\gamma \rangle^\gamma = \frac{V_{\gamma\sigma}}{V_\gamma} \langle \psi_\gamma \rangle \tag{S13}$$

$$\varepsilon_\gamma = \frac{V_\gamma}{V_{\gamma\sigma}} \tag{S14}$$

$$\langle \psi_\gamma \rangle = \varepsilon_\gamma \langle \psi_\gamma \rangle^\gamma \tag{S15}$$

$$\psi_\gamma = \langle \psi_\gamma \rangle^\gamma + \widehat{\psi}_\gamma \tag{S16}$$

$$\left\langle \frac{\partial \psi_\gamma}{\partial t} \right\rangle = \frac{\partial \langle \psi_\gamma \rangle}{\partial t} - \frac{1}{V_{\gamma\sigma}} \int \psi_\gamma \, u_{\gamma\sigma} \cdot \mathbf{n}_{\gamma\sigma} dS \tag{S17}$$

$$\langle \nabla \psi_\gamma \rangle = \nabla \langle \psi_\gamma \rangle + \frac{1}{V_{\gamma\sigma}} \int \psi_\gamma \, \mathbf{n}_{\gamma\sigma} dS \tag{S18}$$

Where, V is representative volume of porous solids with its neighborhood electrolyte liquid, $\psi$ is a quantity. The formulas and theorems represented in equations (S9)-(S18) above provide a theoretical basis to switch from an average value for the derivative of a quantity to the derivative of an average value for a quantity. Those formulas and theorems from related previous studies develop a general framework upon which to build a macroscopic mathematical model that can capture the flow dynamics through the porous electrode. The mathematical model for describing flow through the porous electrode is achieved by averaging mass conservation or continuity and averaging Navier-Stokes' equation

$$\nabla \cdot u_f = 0 \tag{S19}$$

$$0 = -\varepsilon_\gamma \nabla \langle p_\gamma \rangle^\gamma + \mu_\gamma \nabla^2 \langle \mathbf{u}_\gamma \rangle - \mu_\gamma \varepsilon_\gamma \mathbf{k}_{\gamma\sigma}^{-1} \langle \mathbf{u}_\gamma \rangle + \varepsilon_\gamma \rho_\gamma \mathbf{g} \tag{S20}$$

Eq. (S20) is the macroscopic mathematical model for the flow dynamic motion in the porous electrode. $\varepsilon_\gamma$, $\mu_\gamma$, $\rho_\gamma$, $\mathbf{k}_{\gamma\sigma}$, $\langle p_\gamma \rangle^\gamma$ and $\langle \mathbf{u}_\gamma \rangle$ are replaced with $\varepsilon$, $\mu$, $\rho$, $k$, $\langle p_p \rangle$, and $\langle u_p \rangle$ or $\langle v_p \rangle$ or $\langle \omega_p \rangle$, then the Eqs. (S19) and (S20) can be rewritten along the X, Y and Z directions

$$\frac{\partial \langle u_p \rangle}{\partial X} + \frac{\partial \langle v_p \rangle}{\partial Y} + \frac{\partial \langle w_p \rangle}{\partial Z} = 0 \tag{S21}$$

$$0 = -\varepsilon \frac{\partial \langle p_p \rangle}{\partial X} + \mu \left( \frac{\partial^2 \langle u_p \rangle}{\partial X^2} + \frac{\partial^2 \langle u_p \rangle}{\partial Y^2} + \frac{\partial^2 \langle u_p \rangle}{\partial Z^2} \right) - \mu \varepsilon k^{-1} \langle u_p \rangle \tag{S22}$$

$$0 = -\varepsilon \frac{\partial \langle p_p \rangle}{\partial Y} + \mu \left( \frac{\partial^2 \langle v_p \rangle}{\partial X^2} + \frac{\partial^2 \langle v_p \rangle}{\partial Y^2} + \frac{\partial^2 \langle v_p \rangle}{\partial Z^2} \right) - \mu \varepsilon k^{-1} \langle v_p \rangle \tag{S23}$$

$$0 = -\varepsilon \frac{\partial \langle p_p \rangle}{\partial Z} + \mu \left( \frac{\partial^2 \langle w_p \rangle}{\partial X^2} + \frac{\partial^2 \langle w_p \rangle}{\partial Y^2} + \frac{\partial^2 \langle w_p \rangle}{\partial Z^2} \right) - \mu \varepsilon k^{-1} \langle w_p \rangle - \varepsilon \rho g \tag{S24}$$

*Maximum Current Density Model*

The concept of "maximum current density" model proposed is based on the electrolyte mass balance through the porous electrode and Faraday's law of electrolysis[S20]

$$m = \left(\frac{q}{F}\right)\left(\frac{M}{n}\right) \tag{S25}$$

Where, m is the mass of electrolyte flow reactants through the porous electrode, q is the total electric charge, F is the Faraday's constant, M is the molar mass of electrolyte flow reactants and n is the number of electrons transferred in the chemical reactions. The relation of total electric charge with current and time yields

$$q = It \tag{S26}$$

Where, I is current, t is reaction time. Equations (S25) and (S26) give

$$m = \left(\frac{It}{F}\right)\left(\frac{M}{n}\right) \tag{S27}$$

$$\dot{m} = \left(\frac{It}{nF}\right) \tag{S28}$$

Where, $\dot{m} = m/M$

The molar mass flow rate now can be written as

$$\dot{m} = \left(\frac{I}{nF}\right) \tag{S29}$$

Where, $\dot{m} = m/t$

Based on the proposed concept that the total ions are consumed as the electrolyte flow reactants that penetrate through the interface between the serpentine flow channel and porous electrode into the porous electrode and all reactants are consumed, then

$$\dot{m} = (Q_p)_{tot} c_{in} \tag{S30}$$

Where, $(Q_p)_{tot}$ is total electrolyte flow reactants through the porous electrode. $(Q_p)_{tot}$ is related to the geometry of the flow cell (e.g. length, width and thickness of flow passage, landing channels and porous electrode), properties of electrolyte (e.g. density and dynamic viscosity), properties of porous electrode (e.g. porosity and permeability), initial ion concentration and entrance volumetric flow rate (or inlet/entrance mean linear velocity). Then, the maximum current is derived by Eqs. (S29) and (S30)

$$I_m = nFc_{in}(Q_p)_{tot} \tag{S31}$$

Based on the contact area between the porous electrode and ion selective membrane, the mathematical model of maximum current density associated with total flow penetration is obtained

$$i_{max} = \frac{nFc_{in}(Q_p)_{tot}}{A} \tag{S32}$$

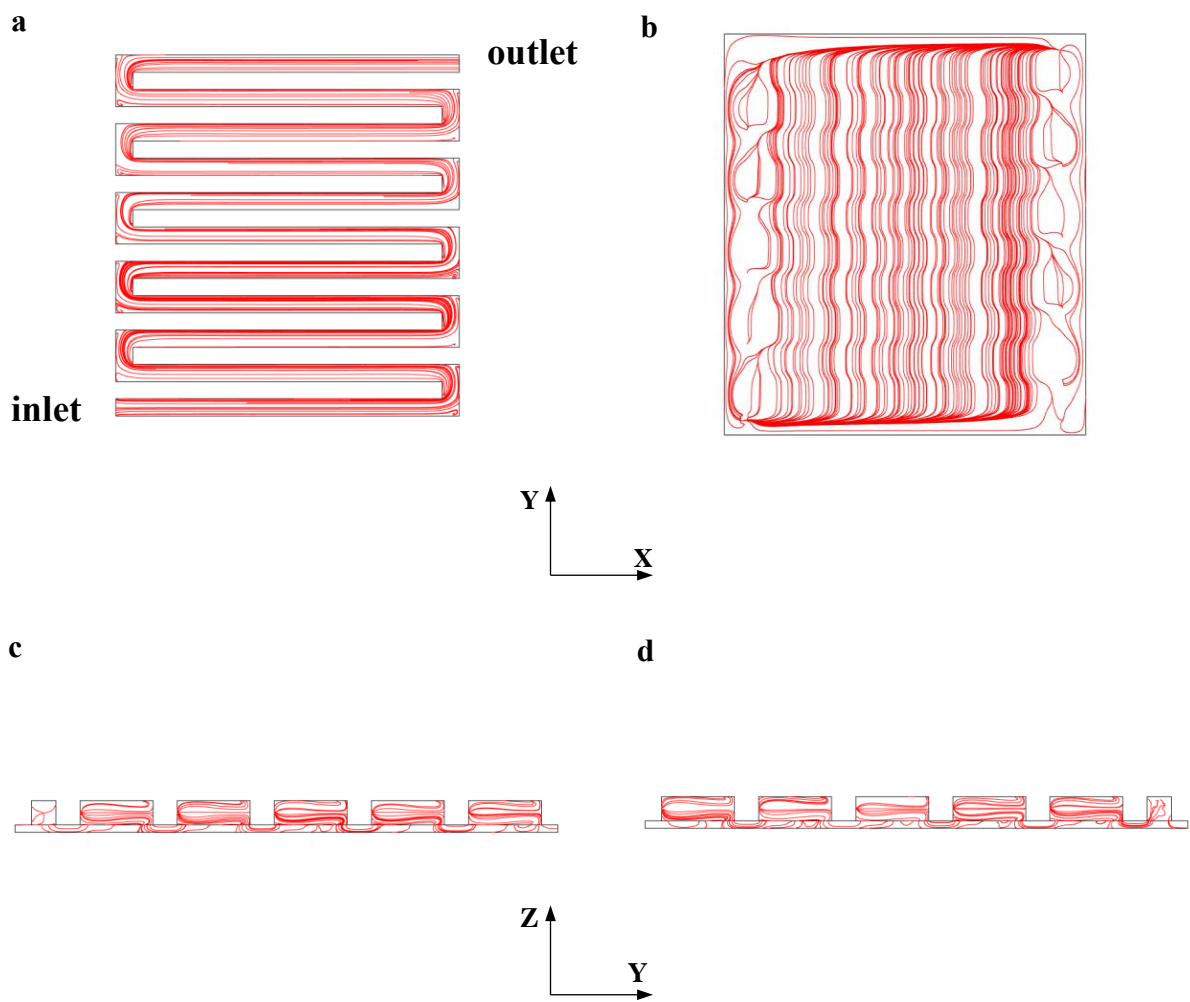

**Figure S3: Electrotype flow streamlines. a,** streamline: (u, v) in the XY plane, Z=$t_p$ (middle thickness of the serpentine flow channel). **b,** streamline: (u, v) in the XY plane, Z=$t_p+t_{fp}/2$ (middle thickness of the porous electrode). **c,** streamline: (v, w) in the YZ plane, X=$(w_p-w_{fp}+w_{cc})/2$ (middle width of the corner channels that are on the inlet side). **d,** streamline: (v, w) in the YZ plane, X=$(w_p+w_{fp}-w_{cc})/2$ (middle width of the corner channels that are on the outlet side).

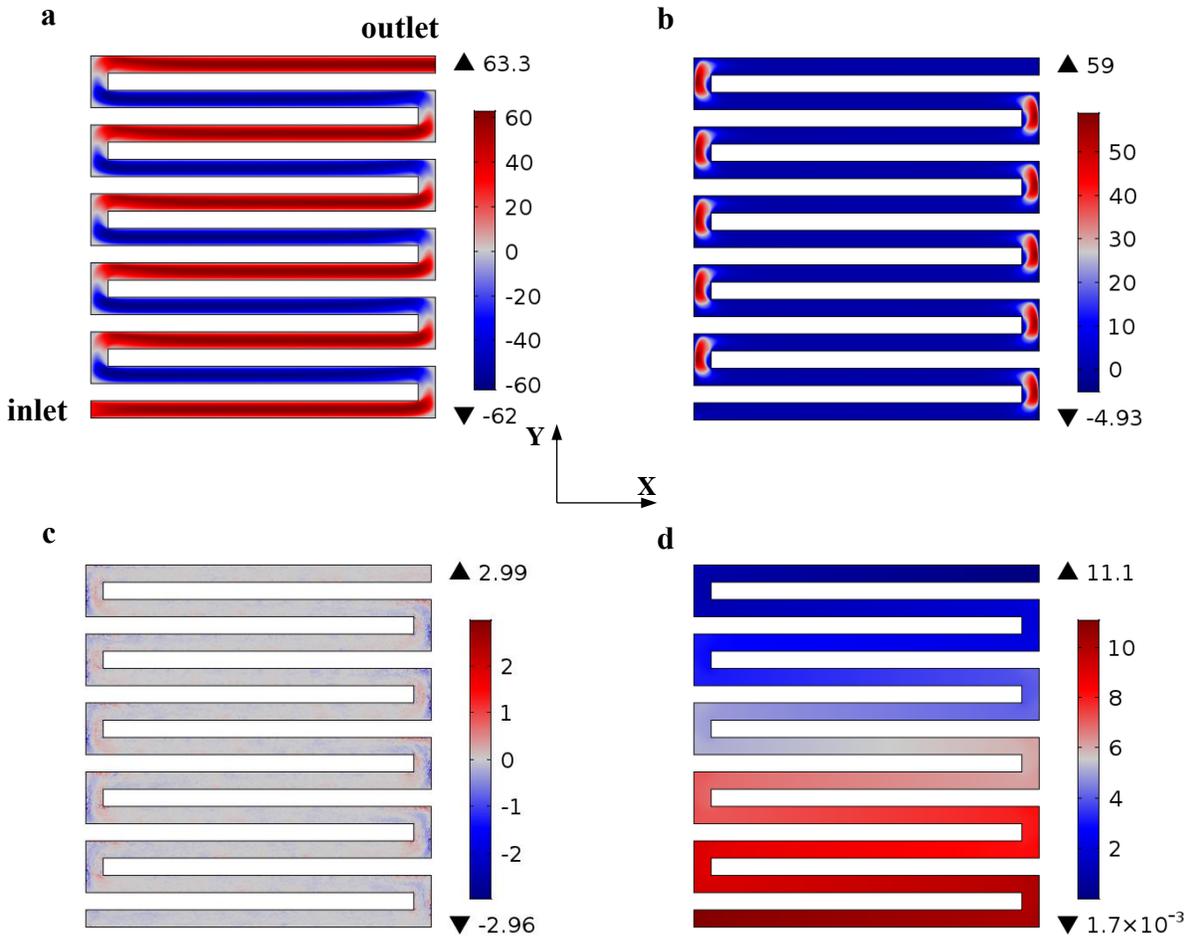

**Figure S4: Flow velocity and pressure distributions at middle thickness of the serpentine flow channel, $Z=t_p+t_{fp}/2$. a,** u distribution, X-component in the XY plane, unit: cm s$^{-1}$. **b,** v distribution, Y-component in the XY plane, unit: cm s$^{-1}$. **c,** w distribution, Z-component in the XY plane, unit: cm s$^{-1}$. **d,** pressure distribution in the XY plane, unit: kPa.

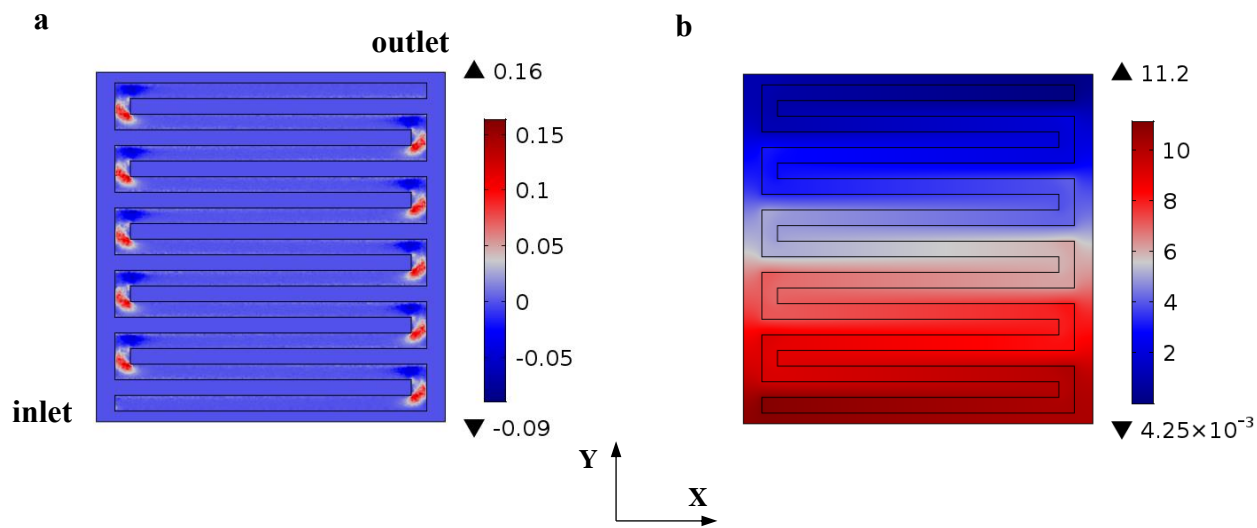

**Figure S5: Flow velocity and pressure distributions at interface between the serpentine flow channel and porous electrode (single layer of SGL 10AA carbon paper electrode), Z=$t_p$.** **a,** v distribution, Y-component in the XY plane, unit: cm s$^{-1}$. **b,** pressure distribution in the XY plane, unit: kPa.

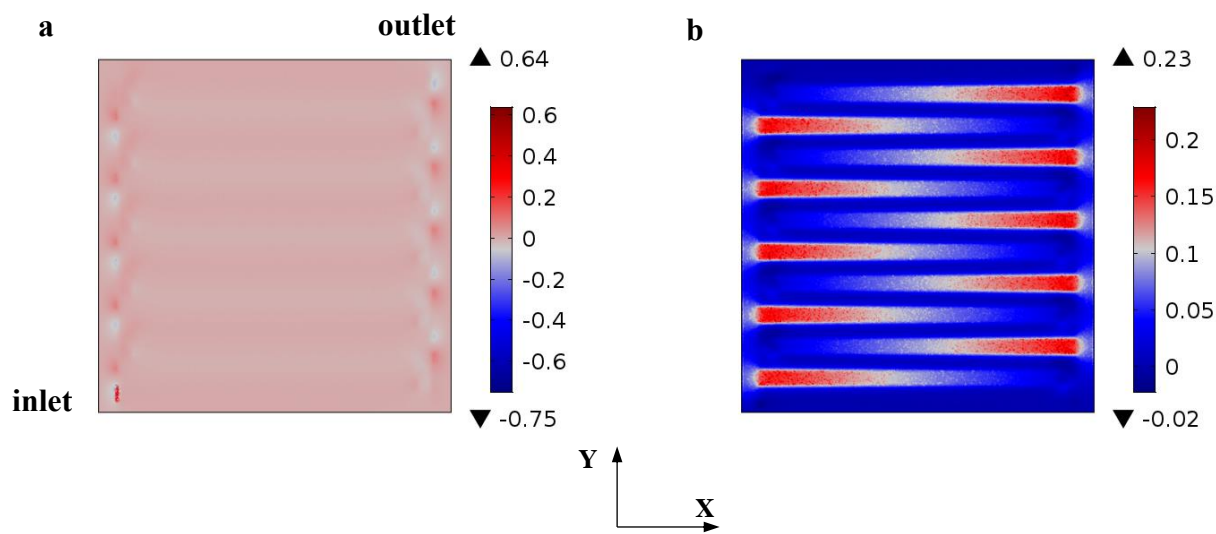

**Figure S6: Flow velocity distributions at middle thickness of the porous electrode (single layer of SGL 10AA carbon paper electrode). a,** u distribution, X-component in the XY plane, unit: cm s$^{-1}$. **b,** v distribution, Y-component in the XY plane, unit: cm s$^{-1}$.

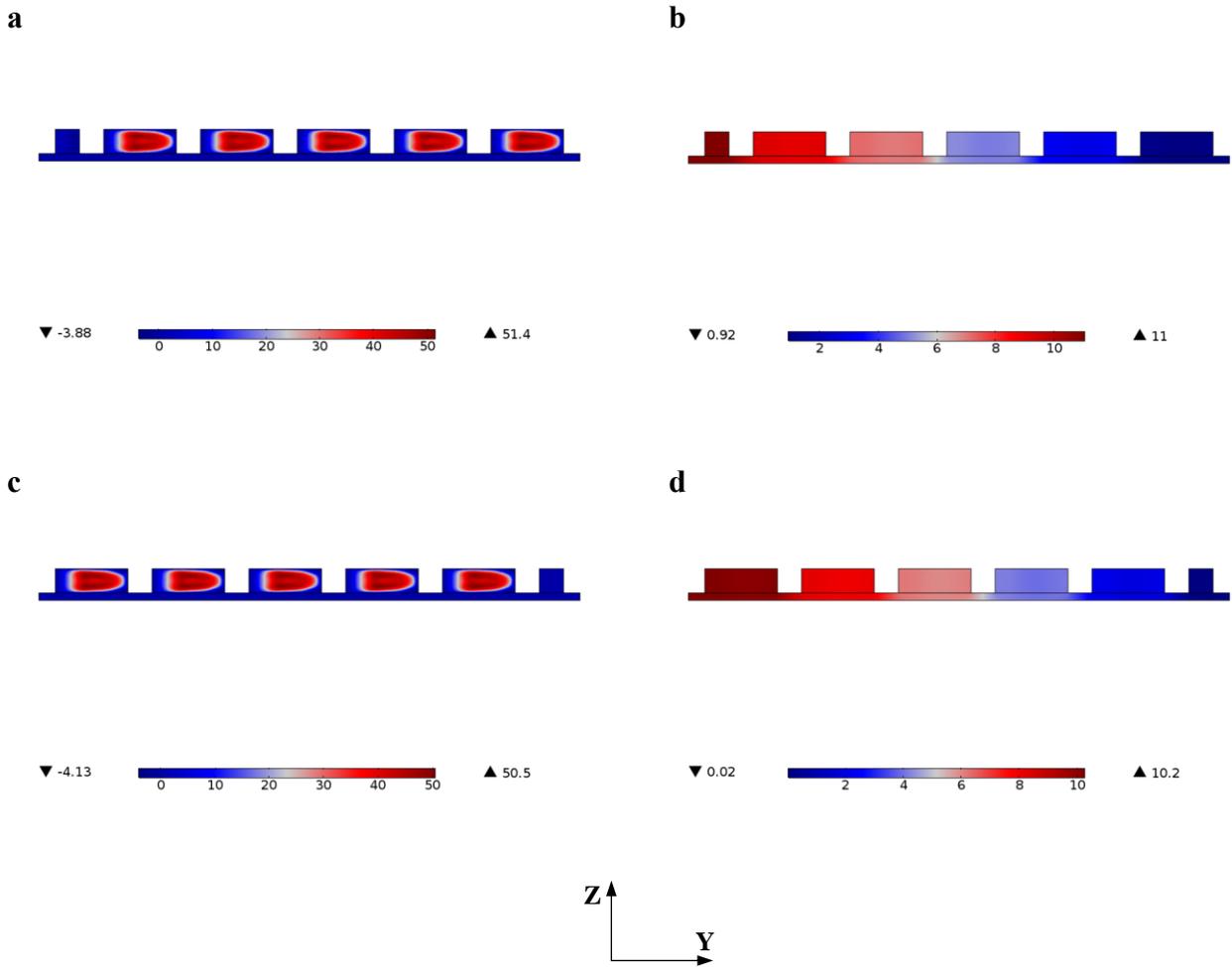

**Figure S7: Flow velocity and pressure distributions at YZ planes, single layer of SGL 10AA carbon paper electrode. a,** (v, w) distribution in the YZ plane, $X=(w_p-w_{fp}+w_{lc})/2$ (middle width of landing channels that are on the inlet side). **b,** pressure distribution in the YZ plane, $X=(w_p-w_{fp}+w_{lc})/2$ (middle width of landing channels that are on the inlet side). **c,** (v, w) distribution in the YZ plane, $X=(w_p+w_{fp}-w_{lc})/2$ (middle width of landing channels that are on the outlet side). **d,** pressure distribution in the YZ plane, $X=(w_p+w_{fp}-w_{lc})/2$ (middle width of landing channels that are on the outlet side).

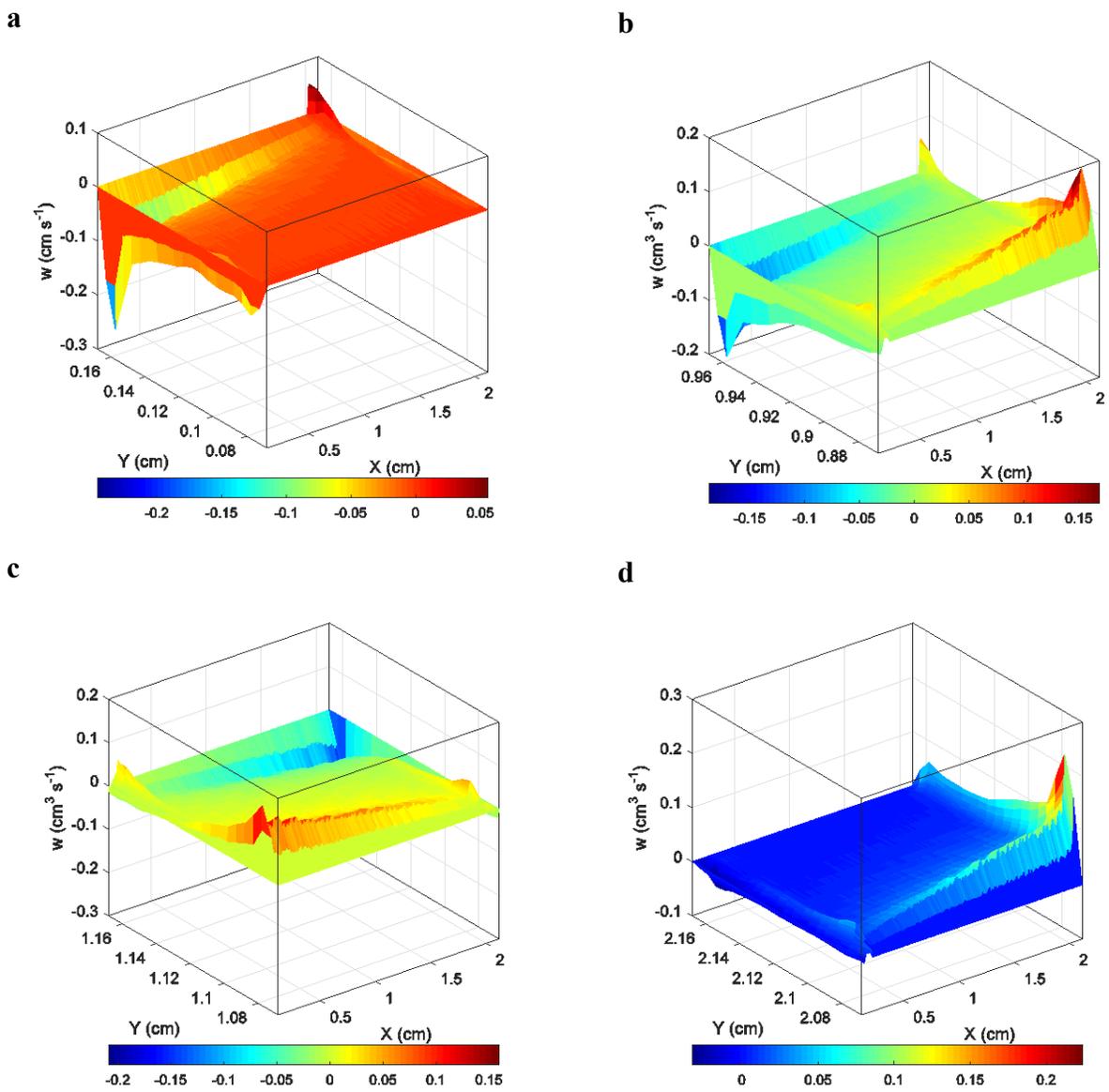

**Figure S8: Volumetric flow penetration w velocity contours along the +Z and −Z directions at the interfaces between the flow passages and porous electrode**. **a,** $\Omega_1$. **b,** $\Omega_9$. **c,** $\Omega_{11}$. **d,** $\Omega_{21}$, unit: cm s$^{-1}$.

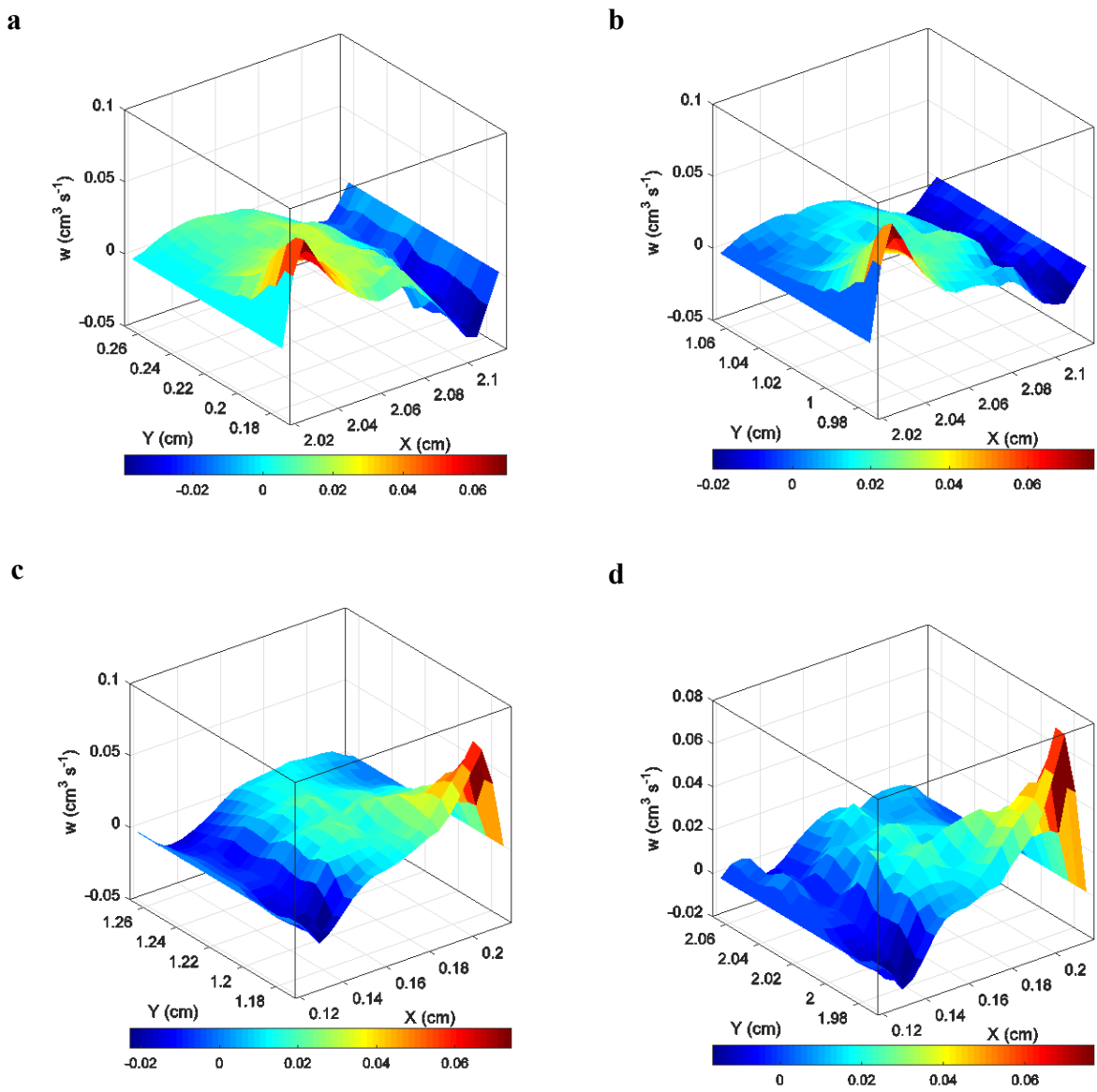

**Figure S9: Volumetric flow penetration w velocity contours along the +Z and –Z directions at the interfaces between the flow passages and porous electrode**. **a,** $\Omega_2$. **b,** $\Omega_{10}$. **c,** $\Omega_{12}$. **d,** $\Omega_{20}$, unit: cm s$^{-1}$.

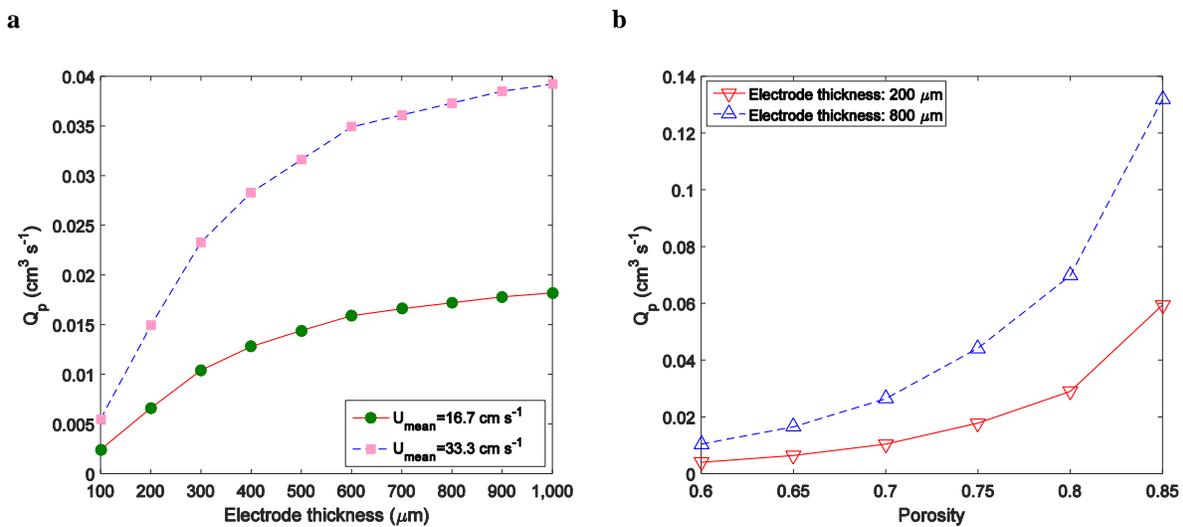

**Figure S10: Factor effects on volumetric flow penetration into the porous electrode. a,** electrode thickness after compression, ranging from 100 μm to 1,000 μm. **b,** porosity after compression, ranging from 0.6 to 0.85.

## Section C: References